\documentclass[aps,eqsecnum,preprintnumbers]{revtex4}
\usepackage{axodraw}
\newcommand{\beq}{\begin{equation}}
\newcommand{\eeq}{\end{equation}}
\newcommand{\beqa}{\begin{eqnarray}}
\newcommand{\eeqa}{\end{eqnarray}}

\newcommand{\lslash}[1]{#1\llap/}

\newcommand{\Eq}[1]{Eq.\ (\ref{#1})}
\newcommand{\Eqs}[2]{Eqs.\ (\ref{#1}) and (\ref{#2})}

\begin{document}

\preprint{hep-ph/0309240}
\title{
The electromagnetic vertex of neutrinos in an electron
background and a magnetic field
}
\author{Jos\'e F. Nieves}
\affiliation{Laboratory of Theoretical Physics, 
Department of Physics, P.O. Box 23343\\
University of Puerto Rico, 
R\'{\i}o Piedras, Puerto Rico 00931-3343}

\date{September 2003}

\begin{abstract}

We study the electromagnetic vertex function of a neutrino that
propagates in an electron background in the
presence of a static magnetic field. The
structure of the vertex function under the stated conditions is determined
and it is written down in terms of a minimal and complete set of tensors.
The one-loop expressions for all the form factors is given,
up to terms that are linear in the magnetic field, 
and the approximate integral formulas that hold
in the long wavelength limit are obtained. We
discuss the physical interpretation of some of the form factors
and their relation with the
concept of the \emph{neutrino induced charge}.
The neutrino acquires a \emph{longitudinal} and a \emph{transverse} charge,
due to the fact that the form factors depend
on the transverse and longitudinal components of the photon
momentum independently.
We compute those form factors
explicitly in various limiting cases and find
that the longitudinal and transverse charge
are the same for the case of a non-relativistic electron gas, 
but not otherwise. 

\end{abstract}

\maketitle

%
% sec 1
%
\section{Introduction and Summary}

In the study of the electromagnetic properties 
of neutrinos in a medium, the two quantities of particular interest
are the neutrino self-energy in the presence of an external
magnetic field, and the neutrino electromagnetic vertex function.
The former quantity determines the neutrino dispersion relation
as it propagates in a magnetized medium, while the latter
determines the relevant couplings in the calculation of the neutrino
electromagnetic processes that occur in the medium, such as
plasmon decay and Cherenkov radiation, among others. 
The electromagnetic
couplings of the neutrino can be viewed in terms of
effective dipole moments and an electric charge induced by the 
medium \cite{np1}.
Although such identifications reveal some of the properties of the
neutrino, both the intrinsic ones as well as those that result from its 
interaction with matter, for
the practical calculation of the transition rates of the various
processes the full electromagnetic vertex function is
needed.
Since the original calculations of those quantities more
than a decade ago \cite{oraevsky,sawyer,dnp1},
some of their possible physical effects
have been considered over the last few years 
\cite{dnp2,semikoz,semikozE,nunokawa,
dn1,dnp3,kicks1,kicks2,kicks3,kicks4,kicks5}.
In addition, the calculations have been improved and refined using a variety 
of methods and techniques \cite{esposito,kimetal,elmfors,dn2}.

Recently, there has been interest on another closely
related quantity, namely the electromagnetic couplings
of the neutrino when it propagates in a medium in the presence of
a magnetic field.
The main issue is here is to determine how
the neutrino electromagnetic processes that take place in a medium, but
in the absence of a magnetic field, are modified when a magnetic
field is present. From a technical point of view, the relevant
quantity is the neutrino electromagnetic vertex function, which
must be calculated taking the presence of the magnetic field into account.

This problem has been approached from two points of view. On 
one hand, the vertex function has been calculated taking the
magnetic field into account, but neglecting the effects of the matter
medium\ \cite{raffelt}. 
On the other hand, the effect of both the magnetic field and
the medium have been taken into account\ \cite{ganguly}, but only
for calculating the coupling that
was identified as the induced neutrino charge, and not
for the calculation of the full vertex function. 
Thus there does not exist a complete 
calculation of the full vertex function that 
takes into account the simultaneous presence of the magnetic field
and the background medium.

In the this work we have taken precisely this problem.
Our goal is to calculate the complete one-loop electromagnetic
vertex function, including the effects of the medium
and the linear terms in the magnetic field. To this end,
we determine the most general form of the neutrino vertex function
in the physical environment and under the conditions that we are considering. 
The vertex function is
written in terms of a complete and minimal set of tensors,
that are consistent with the requirements that follow from the chiral
nature of the neutrino interactions and electromagnetic gauge
invariance. Such a decomposition is generally useful in its
own right, and it is also a useful reference point
for the actual computation of the vertex function. Thus,
we give the formulas for all the corresponding
form factors in terms of momentum integrals over the 
background electron distribution functions. 
For arbitrary values of the photon momentum and/or general
distributions functions the integrals are normally not doable,
but they can be evaluated for various limiting cases and approximations that
correspond to realistic situations. A particularly useful one
that we consider is the so-called \emph{long wavelength limit}, which is valid
when the photon energy and momentum are smaller
than the typical energy of an electron in the gas. 
In this limit the integrals that appear in the formulas for the form factors
simplify but yet they remain valid for general forms of the momentum
distributions, and for a wide range of values of the photon energy
and momentum subject only to the restriction stated above.
The simplified integral expressions so obtained can be used
to compute explicitly the form factors for different kinematical
regimes, such as the static limit (taking the photon energy to zero),
and for various conditions of the background such as the classical
or degenerate gas. 

We observe that, apart from the photon energy,
the form factors depend separately on the perpendicular and 
parallel components of the photon momentum.
That is to say, if we denote those three quantities by
$\omega,Q_\perp,Q_\parallel$ respectively, the form factors depend
on those three variables and not just on $\omega$ and 
$Q^2 = Q^2_\perp + Q^2_\parallel$ as it is the case in the absence
of the magnetic field. A consequence of this is that,
in contrast to the latter case, there is no
unique meaning to the concept of the induced neutrino charge,
or in fact any other electromagnetic moment,
when the magnetic field is non-zero. Briefly, the reason is that 
the zero (photon) momentum limit of the 
static form factors (i.e., evaluated for $\omega = 0$) can be
taken in two different ways, according to whether we set 
$Q_\perp = 0$ first and $Q_\parallel\rightarrow 0$ afterwards
or the other way around. If we insist on identifying a quantity such
as the induced charge by looking at the static vertex function in the
zero (photon) momentum limit, we are thus forced to define two different 
quantities, which in the specific case of the neutrino induced charge
we denote by $e^\parallel_\nu$ and $e^\perp_\nu$. 
We stress that the form factors have a different
value in the two limits, and this is not a mathematical ambiguity
but actually a reflection of the fact that, in the presence of
the magnetic field, the two ways of taking the zero 
momentum limit correspond to different physical situations.
%
%\footnote{This is analogous to taking the photon self-energy $\pi(\omega,Q)$,
%and look at its value in the two limits $\pi(0,Q\rightarrow 0)$
%and $\pi(\omega\rightarrow 0,0)$. The first gives the Debye
%screening length while the second one gives the plasma frequency.}
%
As an illustration, and to further clarify some of these issues,
we compute explicitly
the form factors that are related to $e^\parallel_\nu$ and $e^\perp_\nu$.
As we show, $e^\parallel_\nu$ and $e^\perp_\nu$ have the same value
for the case of a non-relativistic electron gas, but they are
given by different formulas otherwise.
In order to establish contact
with previous calculations\ \cite{ganguly}, 
the quantity calculated there is identified with what we call 
$e^\parallel_\nu$. However, some differences in the results are found
and are noted.

In what follows we present the details of the calculations,
divided as follows.
An essential ingredient is the \emph{linearized form}
of the thermal propagator for an electron in a magnetic field.
In Section\ \ref{sec:linear},
the linearized form is obtained from the Schwinger formula for
the electron propagator, generalized to include the effects
of the background medium, and approximating it up to terms that are
linear in the magnetic field. 
In Section\ \ref{sec:calculation}, the linearized propagator is used
to obtain the one-loop formula for the neutrino electromagnetic vertex 
up to terms that are linear in the magnetic field.
In Section \ref{sec:structure} we determine the structure
of the vertex function and decompose it in terms of a minimal
set of tensors. The end result of this procedure is
a set of formulas for the corresponding form factors, expressed
as integrals over the
momentum distribution functions of the electrons in the medium.
In Section\ \ref{sec:approximate}, the approximate formulas that
are valid in the long wavelength limit are derived first, and they are then 
used to evaluate explicitly the form factors
that are related to $e^\parallel_\nu$ and $e^\perp_\nu$
for various specific cases.
Section\ \ref{sec:conclusions} contains our conclusions.
In addition there are two appendices.
The general decomposition of the vertex function is guided by
several requirements, such as the transversality condition which
of course is a general consequence of the electromagnetic gauge invariance.
However, it is neither obvious to see nor trivial to prove, that
the one-loop expression for the vertex function satisfies the
transversality condition. The proof that it in fact does so
is supplied in those appendices.

%
% sec 2
%
\section{The linearized electron propagator}
\label{sec:linear}
In the notation of Ref.\ \cite{dns}, the Schwinger propagator for
the electron in a magnetic field can be written in the form
\beq
\label{SF}
iS_F(p) = \int_0^\infty\,d\tau G(p,s) e^{i\tau\Phi(p,s) - \epsilon\tau} \,,
\eeq
where the variable $s$ is defined by
\beq
s = |e|B\tau \,,
\eeq
and the exact formulas for $G(p,s)$ and $\Phi(p,s)$ are given
in Eq.\ (3.4) of that reference. The linear approximation that
we propose to use consists in retaining up to terms that are linear in $B$.
To this order,
\beqa
\label{PhiG}
\Phi & = & p^2 - m^2 \nonumber\\
G & = & \lslash{p} + m + i\tau |e|B G_B \,,
\eeqa
where
\beq
\label{GB}
G_B = \gamma_5\left[(p\cdot b)\lslash{u}
- (p\cdot u)\lslash{b} + m\lslash{u}\lslash{b}\right] \,,
\eeq
so that
\beq
\label{SFsplit}
S_F = S_0 + S_B \,,
\eeq
where
\beqa
\label{S0SB}
iS_0 & = & i\frac{\lslash{p} + m}{p^2 - m^2}\nonumber\\
iS_B & = & \frac{-i}{(p^2 - m^2)^2}(|e|BG_B)\,.
\eeqa
The vector $u^\mu$ that appears in \Eq{GB} is the velocity four-vector
of the medium which, in the frame of reference in which the medium is at rest, 
takes the form
\beq
u^\mu = (1,0) \,.
\eeq
We have also introduced the vector $b^\mu$ which is such that, in that frame,
\beq
b^\mu = (0,\hat b) \,,
\eeq
where the magnetic field is given by
\beq
\vec B = B\hat b \,.
\eeq

The thermal electron propagator which, in addition to the magnetic
field, incorporates the effect of the electron background,
is given by
\beq
\label{Sethermal}
S_e = S_F - [S_F - \bar S_F]\eta_e \,,
\eeq
where
\beq 
\label{etae} 
\eta_e(p\cdot u) = \theta(p\cdot u)f_e(p\cdot u) +
\theta(-p\cdot u)f_{\bar e}(-p\cdot u)\,, 
\eeq
with
\beqa 
f_e(x) & = & \frac{1}{e^{\beta(x - \mu_e)} + 1} \nonumber\\
f_{\bar e}(x) & = & \frac{1}{e^{\beta(x + \mu_e)} + 1} \,. 
\eeqa
Here $\beta$ stands for the inverse temperature and $\mu_e$ the
electron chemical potential. Using Eqs.\ (\ref{SFsplit}),
(\ref{S0SB}), (\ref{GB}), the complete propagator can be expressed as
\beq
\label{Seexpanded}
S_e = S_0 + S_B + S_T + S_{TB} \,,
\eeq
where 
\beqa
\label{STB}
iS_T & = & -2\pi\delta(p^2 - m^2)\eta_e (\lslash{p} + m)\nonumber\\
iS_{TB} & = & -2\pi\delta^\prime(p^2 - m^2)\eta_e (|e|BG_B) \,,
\eeqa
where $G_B$ is given in \Eq{GB} and
the notation $\delta^\prime$ denotes the derivative of the
delta function with respect to its argument. 

Although we will not give the details here, it is reassuring to mention
that when this linearized propagator is used to compute
the self-energy diagrams for the neutrino in an electron gas,
the usual expression for the magnetic field contribution to the
neutrino dispersion relation, or equivalently the index refraction, 
is reproduced.

%
% sec 3
%
\section{Calculation of the vertex function}
\label{sec:calculation}
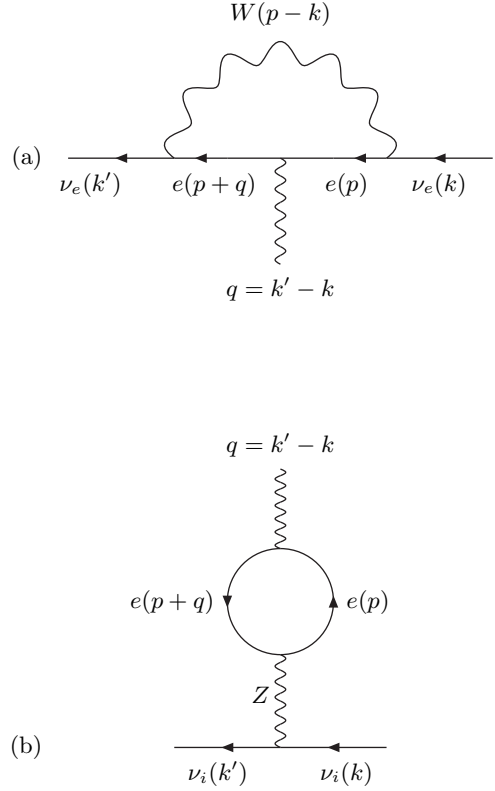
\begin{figure}
\begin{center}
%
% Exchange diagram
%
\begin{picture}(180,200)(-90,-100)
%\Text(0,-80)[cb]{(a)}
\Text(-90,0)[cr]{(a)}
\Photon(0,0)(0,-40){2}{6}
\Text(0,-45)[ct]{$q = k' - k$}
\ArrowLine(80,0)(40,0)
\ArrowLine(40,0)(20,0)
\Text(60,-10)[c]{$\nu_e(k)$}
\ArrowLine(-20,0)(-40,0)
\Line(20,0)(-20,0)
\Text(25,-10)[c]{$e(p)$}
\Text(-25,-10)[c]{$e(p + q)$}
\ArrowLine(-40,0)(-80,0)
\Text(-60,-10)[cr]{$\nu_e(k')$}
\PhotonArc(0,0)(40,0,180){4}{6.5}
\Text(0,50)[cb]{$W(p - k)$}
\end{picture}
\\[12pt]
%
% Tadpole diagram
%
%\begin{picture}(100,100)(-50,-30)
\begin{picture}(100,130)(-50,-20)
%\Text(0,-80)[cb]{(b)}
\Text(-90,0)[rc]{(b)}
\Photon(0,75)(0,105){2}{6}
\Text(0,110)[bc]{$q = k' - k$}
\ArrowLine(40,0)(0,0)
\Text(35,-10)[cr]{$\nu_i(k)$}
\ArrowLine(0,0)(-40,0)
\Text(-35,-10)[cl]{$\nu_i(k')$}
\Photon(0,0)(0,35){2}{6}
\Text(-4,20)[r]{$Z$}
\ArrowArc(0,55)(20,-90,90)
\ArrowArc(0,55)(20,90,270)
\Text(25,55)[l]{$e(p)$}
\Text(-25,55)[r]{$e(p + q)$}
\end{picture}
\caption[]{One-loop diagrams for the neutrino electromagnetic
vertex in an electron background, to order $1/M^2_W$.
\label{fig:nuemvertex}}
\end{center}
\end{figure}

The off-shell neutrino electromagnetic vertex function $\Gamma^\mu$
is defined such that the matrix element of the electromagnetic current
between neutrino states with momenta $k$ and $k^\prime$ is given
by
\beq
\langle k^\prime|j_\mu(0)|k\rangle =
\bar u(k^\prime)\Gamma_\mu u(k) \,.
\eeq
To order $1/M^2_W$, the one-loop diagrams that are relevant to the 
calculation of $\Gamma^\mu$ are shown in Fig.\ \ref{fig:nuemvertex}. 
Their contributions are 
\beqa 
-i\Gamma^{(W)}_{\nu}(q) & = &
- \left(\frac{eg^2}{2 M^2_W}\right)
\int\frac{d^4p}{(2\pi)^4}\gamma^\mu
L iS_e(p + q)\gamma_\nu iS_e(p)\gamma_\mu L\,,\nonumber\\
-i\Gamma^{(Z)}_{\nu}(q)& = & \left(\frac{eg^2}
{4\cos^2\theta_W M^2_Z}\right) \gamma^\mu L
\int\frac{d^4p}{(2\pi)^4}\mbox{Tr}\,[iS_e(p + q)\gamma_\nu
iS_e(p)\gamma_\mu (a_e + b_e \gamma_5)]\,, 
\eeqa
where
\beq
q = k^\prime - k
\eeq
is the  momentum of  the incoming photon,
$L  =  \frac{1}{2}(1 - \gamma_5)$ and, in the standard model
\beqa
a_e & = & - \frac{1}{2} +2 {\rm sin}^2\theta_W\,,\\
b_e & = & \frac{1}{2}\,.
\eeqa

Using the identity
\beq 
\gamma^\mu L M \gamma_\mu L = -\mbox{Tr}\,(M\gamma_\mu L)
\gamma^\mu L\,, 
\eeq
which holds for any $4\times4$ matrix $M$, the charge current contribution
can be put in the same form as the neutral current one, namely  
\beq 
-i\Gamma^{(W)}_{\nu}(q) = \left(\frac{eg^2}{2M^2_W}\right)
\gamma^\mu L \int\frac{d^4p}{(2\pi)^4}\mbox{Tr}\,[iS_e(p + q)
\gamma_\nu iS_e(p)\gamma_\mu L] \,. 
\eeq
When Eq.\ (\ref{Seexpanded}) is substituted in 
the above expressions for $\Gamma^{(W,Z)}_\nu$, several terms 
are generated and we want to retain only those that, schematically, involve
the products $S_{TB}S_0$ or $S_T S_B$, i.e., the terms that depend
both on $B$ and the temperature. Proceeding
in this way, we then have
\beqa 
\label{TVAdef}
\Gamma^{(W)}_{\nu} & = & -\left(\frac{eg^2}{4M^2_W}\right) 
\gamma^\mu L \left(T^{(V)}_{\mu\nu} + T^{(A)}_{\mu\nu}\right)\,,\nonumber\\
\Gamma^{(Z)}_{\nu} & = & -\left(\frac{eg^2}{4M^2_W}\right) 
\gamma^\mu L \left(a_e T^{(V)}_{\mu\nu} - b_eT^{(A)}_{\mu\nu}\right)\,,
\eeqa
where
\beqa
\label{TV}
T^{(V)}_{\mu\nu} & = & i\int\frac{d^4p}{(2\pi)^4}\mbox{Tr}\,
\gamma_\mu \left[ S_0(p')\gamma_\nu S_{TB}(p) +
S_{TB}(p')\gamma_\nu S_{0}(p)\right. \nonumber\\
&&\mbox{} + \left.
S_T(p')\gamma_\nu S_{B}(p) +
S_B(p')\gamma_\nu S_{T}(p)\right]\,,\\[12pt]
\label{TA}
T^{(A)}_{\mu\nu} & = & i\int\frac{d^4p}{(2\pi)^4}\mbox{Tr}
\gamma_5\gamma_\mu\left[ S_0(p')\gamma_\nu S_{TB}(p)+
S_{TB}(p')\gamma_\nu S_{0}(p) \right. \nonumber\\
&&\mbox{} + \left.
S_T(p')\gamma_\nu S_{B}(p) +
S_B(p')\gamma_\nu S_{T}(p)\right] \,,
\eeqa
and we have defined
\beq
p' = p + q \,.
\eeq
The vertex function for the electron neutrino is
$\Gamma^{(W)}_\nu + \Gamma^{(Z)}_\nu$, while for
the other neutrino flavors it is just $\Gamma^{(Z)}_\nu$.
Denoting a given neutrino flavor by $\ell = e,\mu,\tau$, then
the neutrino vertex function is
\beq
\Gamma^{(\ell)}_\nu = 
-\left(\frac{eg^2}{4M^2_W}\right) 
\gamma^\mu L \left(\chi^{(\ell)}_V T^{(V)}_{\mu\nu} + 
\chi^{(\ell)}_A T^{(A)}_{\mu\nu}\right)\,,
\eeq
where
\beqa
\chi^{(\ell)}_V & = & \delta_{\ell,e} + a_e \,,\nonumber\\
\chi^{(\ell)}_A & = & \delta_{\ell,e} - b_e \,.
\eeqa

All the four terms in \Eq{TV} involve the traces
\beqa
4L^{(1)}_{\mu\nu} & = & \mbox{Tr}\,\gamma_\mu(\lslash p' + m)
\gamma_\nu G_B(p)\,, \nonumber\\
4L^{(2)}_{\mu\nu} & = & \mbox{Tr}\,\gamma_\mu G_B(p')
\gamma_\nu (\lslash p + m)\,,
\eeqa
and analogously in \Eq{TA},
\beqa
\label{Ktraces}
4K^{(1)}_{\mu\nu} & = & \mbox{Tr}\,\gamma_5\gamma_\mu
(\lslash p' + m)\gamma_\nu G_B(p\,,\nonumber\\
4K^{(2)}_{\mu\nu} & = & \mbox{Tr}\,\gamma_5\gamma_\mu G_B(p')
\gamma_\nu
(\lslash p + m)\,.
\eeqa
It is useful to observe that
$L^{(2)}_{\mu\nu} = L^{(1)}_{\nu\mu}(p'\leftrightarrow p)$ and
$K^{(2)}_{\mu\nu} = K^{(1)}_{\mu\nu}(p\leftrightarrow p')$.
Then, by standard Dirac matrix algebra we find
\beqa
\label{L12}
L^{(1)}_{\mu\nu} & = & i\epsilon_{\mu\nu\alpha\beta}\left(
p\cdot u \ p^{\prime\,\alpha}b^\beta -
p\cdot b \ p^{\prime\,\alpha}u^\beta -
m^2 u^\alpha b^\beta\right) \,,\nonumber\\
L^{(2)}_{\mu\nu} & = & i\epsilon_{\mu\nu\alpha\beta}\left(
- p'\cdot u \ p^{\alpha}b^\beta +
p'\cdot b \ p^{\alpha}u^\beta +
m^2 u^\alpha b^\beta\right) \,,\nonumber\\
\label{K12}
K^{(1)}_{\mu\nu} & = &
p\cdot u \left( p'_\mu b_\nu + b_\mu p'_\nu - p'\cdot b \ g_{\mu\nu}  \right)
- p\cdot b \left( p'_\mu u_\nu + u_\mu p'_\nu - p'\cdot u \ g_{\mu\nu}\right) 
- m^2 \left(u_\mu b_\nu - b_\mu u_\nu \right) \,,\nonumber\\
K^{(2)}_{\mu\nu} & = &
p'\cdot u \left( p_\mu b_\nu + b_\mu p_\nu - p\cdot b \ g_{\mu\nu}  \right)
- p'\cdot b \left( p_\mu u_\nu + u_\mu p_\nu - p\cdot u \ g_{\mu\nu} \right) 
- m^2 \left(u_\mu b_\nu - b_\mu u_\nu \right) \,.
\eeqa
Going back to \Eqs{TV}{TA} and using the above formulas for the
traces, we obtain
\beqa
\label{TVAintegral}
T^{(V)}_{\mu\nu} & = & 4|e|B \int\frac{d^4p}{(2\pi)^3} \eta_e(p.u)\left\{
\frac{-L^{(1)}_{\mu\nu}\delta^\prime(p^2 - m^2)}{(p + q)^2 - m^2} +
\frac{L^{(2)}_{\mu\nu}\delta(p^2 - m^2)}{[(p + q)^2 - m^2]^2}
- (q\rightarrow -q)\right\} \,,\nonumber\\
T^{(A)}_{\mu\nu} & = & 4|e|B \int\frac{d^4p}{(2\pi)^3} \eta_e(p.u)\left\{
\frac{-K^{(1)}_{\mu\nu}\delta^\prime(p^2 - m^2)}{(p + q)^2 - m^2} +
\frac{K^{(2)}_{\mu\nu}\delta(p^2 - m^2)}{[(p + q)^2 - m^2]^2}
+ (q\rightarrow -q)\right\} \,.
\eeqa
In arriving at this expression we have used the standard trick
of making the change of variables
$p\rightarrow p - q$ in those terms where the factor $\eta_e(p^\prime)$
appears, together with the properties
\beqa
L^{(1)}_{\mu\nu}(p\rightarrow p - q) & = & -L^{(2)}_{\mu\nu}(q\rightarrow -q)
\,,\nonumber\\
L^{(2)}_{\mu\nu}(p\rightarrow p - q) & = & -L^{(1)}_{\mu\nu}(q\rightarrow -q)
\,,\nonumber\\
K^{(1)}_{\mu\nu}(p\rightarrow p - q) & = & K^{(2)}_{\mu\nu}(q\rightarrow -q)
\,,
\nonumber\\
K^{(2)}_{\mu\nu}(p\rightarrow p - q) & = & K^{(1)}_{\mu\nu}(q\rightarrow -q)
\,,
\eeqa
which can be seen by inspection of \Eq{K12}.

%
% sec 4
%
\section{Structure of the vertex function}
\label{sec:structure}
In the context of the one loop calculation that we are considering, 
the conservation of the electron vector current implies that
the tensors $T^{(V,A)}_{\mu\nu}$ satisfy the transversality relations
\beqa
\label{transversality}
q^\nu T^{(V)}_{\mu\nu} = q^\mu T^{(V)}_{\mu\nu} & = & 0\,,\nonumber\\ 
q^\nu T^{(A)}_{\mu\nu} & = & 0\,.
\eeqa
It is neither immediately evident nor trivial to prove 
that the one-loop formulas 
given in \Eq{TVAintegral} satisfy such relations but,
as we show explicitly in Appendices\ \ref{sec:transversalityTV} and 
\ref{sec:transversalityTA}, they do. Similarly, there we also
verify that, in contrast to the situation without the magnetic field,
in the present case $q^\mu T^{(A)}_{\mu\nu} \not= 0$.

To exploit \Eq{transversality} we introduce the following set of
mutually orthogonal vectors
\beqa
\label{tildeubt}
\tilde u_\mu & = & \tilde g_{\mu\nu} u^\nu\,, \nonumber\\
\tilde b_\mu & = & \tilde g_{\mu\nu} b^\nu -\frac{q\cdot u}{q^2}
\left(q\cdot u \, b_\mu - q\cdot b \, u_\mu \right)\,,\\
\tilde t_\mu & = & 
\epsilon_{\mu\alpha\beta\gamma} u^\alpha b^\beta q^\gamma \,,\nonumber
\eeqa
where
\beq
\tilde g_{\mu\nu} \equiv g_{\mu\nu} - \frac{q_\mu q_\nu}{q^2}\,.
\eeq
These vectors are transverse to $q_\mu$ and satisfy the closure relation
\beq
\frac{\tilde u_\mu \tilde u_\nu}{\tilde u^2} +
\frac{\tilde b_\mu \tilde b_\nu}{\tilde b^2} +
\frac{\tilde t_\mu \tilde t_\nu}{\tilde t^2} = \tilde g_{\mu\nu}\,.
\eeq
It is useful to note that $\tilde b^\mu$ and $\tilde t^\mu$ also satisfy
\beq
\tilde b \cdot u = \tilde t \cdot u = 0 \,.
\eeq

\subsection{The structure of $T^{(V)}_{\mu\nu}$}
The transversality of $T^{(V)}_{\mu\nu}$ with respect to both
indices implies that it can be expanded in terms of the nine bilinear
products that can be formed out of the set of vectors 
$\{\tilde u, \tilde b, \tilde t\}$. Therefore,
we define the following three diagonal tensors
\beqa
Q_{\mu\nu} & = & 
\frac{\tilde u_\mu \tilde u_\nu}{\tilde u^2}\,,\nonumber\\
B_{\mu\nu} & = & 
\frac{\tilde b_\mu \tilde b_\nu}{\tilde b^2}\,,\nonumber\\
R_{\mu\nu} & = & \tilde g_{\mu\nu} - Q_{\mu\nu} - B_{\mu\nu}\,,
\eeqa
the three symmetric combinations
\beqa
S_{1\mu\nu} & = & 
\tilde b_\mu \tilde t_\nu + (\mu\leftrightarrow\nu)\,, \nonumber\\
S_{2\mu\nu} & = & 
\tilde u_\mu \tilde t_\nu + (\mu\leftrightarrow\nu)\,,\nonumber\\
S_{3\mu\nu} & = & 
\tilde u_\mu \tilde b_\nu + (\mu\leftrightarrow\nu) \,,
\eeqa
and the three antisymmetric ones
\beqa
A_{1\mu\nu} & = & 
\tilde b_\mu \tilde t_\nu - (\mu\leftrightarrow\nu) \,, \nonumber\\
A_{2\mu\nu} & = & 
\tilde u_\mu \tilde t_\nu - (\mu\leftrightarrow\nu) \,,\nonumber\\
A_{3\mu\nu} & = & 
\tilde u_\mu \tilde b_\nu - (\mu\leftrightarrow\nu) \,.
\eeqa

$A_{1\mu\nu}$ and $A_{2\mu\nu}$
can be expressed in a more convenient form by using the relations
\beqa
\tilde u_\nu \epsilon_{\mu\alpha\beta\gamma} \tilde u^\alpha \tilde b^\beta
q^\gamma - (\mu\leftrightarrow \nu)  = \tilde u^2\epsilon_{\mu\nu\alpha\beta}
\tilde b^\alpha q^\beta\nonumber\\
\tilde b_\nu \epsilon_{\mu\alpha\beta\gamma} \tilde b^\alpha \tilde u^\beta
q^\gamma - (\mu\leftrightarrow \nu) = \tilde b^2\epsilon_{\mu\nu\alpha\beta}
\tilde u^\alpha q^\beta \,,
\eeqa
which follow from contracting the identity
\beq
\label{relacionpugnetera}
g_{\lambda\nu}\epsilon_{\mu\gamma\alpha\beta} -
g_{\lambda\mu}\epsilon_{\nu\gamma\alpha\beta} +
g_{\lambda\gamma}\epsilon_{\nu\mu\alpha\beta} -
g_{\lambda\alpha}\epsilon_{\nu\mu\gamma\beta} +
g_{\lambda\beta}\epsilon_{\nu\mu\gamma\alpha} = 0
\eeq
with $\tilde u^\lambda \tilde u^\nu \tilde b^\alpha q^\beta$ in one case,
and with  $\tilde b^\lambda \tilde b^\nu \tilde u^\alpha q^\beta$ in
the other.
Therefore, instead of the antisymmetric combinations $A_{1,2}$,
we will use instead
\beqa
\tilde P_{1\mu\nu} & \equiv i\epsilon_{\mu\nu\alpha\beta}
\tilde u^\alpha q^\beta \,,\nonumber\\
\tilde P_{2\mu\nu} & \equiv i\epsilon_{\mu\nu\alpha\beta}
\tilde b^\alpha q^\beta \,.
\eeqa

Two additional properties of the one-loop formula for $T^{(V)}_{\mu\nu}$, 
which follow from inspection of \Eq{TVAintegral},
are: (i) it is antisymmetric, 
and (ii) it transforms as a pseudo-tensor under parity.
Out of the set of nine tensors defined above, only $\tilde P_1$ and
$\tilde P_2$ share both properties. It is then clear that, to this order, 
the structure of $T^{(V)}_{\mu\nu}$ is simply
\beq
\label{TVform1loop}
T^{(V)}_{\mu\nu} =
\tilde T^{(V)}_1 \tilde P_{1\mu\nu} + \tilde T^{(V)}_2 \tilde P_{2\mu\nu} \,.
\eeq
The form factors $T^{(V)}_{1,2}$ are, in general, functions of the
three scalar variables $\omega$, $Q_\parallel$ and $Q_\perp$ which
are defined by
\beqa
\label{kinematicvars}
\omega & = & q\cdot u \,, \nonumber\\
Q_\parallel & = & -q\cdot b\,,\nonumber\\
Q_\perp & = & \sqrt{Q^2 - Q^2_\parallel}\,,
\eeqa
where
\beq
Q^2 = |\vec Q|^2 = \omega^2 - q^2\,.
\eeq
These variables correspond to the photon energy in the rest frame
of the medium and the components of the photon momentum $\vec Q$ that are 
parallel and perpendicular, respectively, to $\vec B$ in this frame.
It is important to stress that in general, besides $\omega$, the form 
factors introduced in \Eq{TVform1loop} depend on the variables $Q_\perp$ 
and $Q_\parallel$, separately. This contrasts with the situation without
the magnetic field, in which case the form factors depend only on 
$\omega$ and $Q$. As we will see in Section\ \ref{subsec:interpretation},
this requires some extra care if we want to interpret the form
factors in terms of the static electromagnetic moments of the neutrino.

The form factors are determined by contracting both sides of 
\Eq{TVAintegral} with the tensors $\tilde P_{1,2}$ and using the 
relations
\beqa
\tilde P_1^2 & = & -2Q^2\,, \nonumber\\
\tilde P_2^2 & = & -2Q^2 \, \frac{Q_\perp^2}{q^2}\,,\\
\tilde P_1\tilde P_2 & = & 0 \,, \nonumber
\eeqa
with $\tilde P_i\tilde P_j= \tilde P_i^{\mu\nu}\tilde P_{j\mu\nu}$.
In this way, we obtain
\beqa
\tilde T_1^{(V)} & = & \frac{-1}{Q^2}\,T_1^{(V)}\,,\nonumber\\
\tilde T_2^{(V)} & = & \frac{1}{Q^2_\perp} \left(
T_2^{(V)} + \frac{\omega Q_\parallel}{Q^2} T_1^{(V)}\right)\,,
\eeqa
where $T_{1,2}^{(V)}$ are the integrals 
\beq
\label{TVform}
T_i^{(V)} = 
4|e|B \int\frac{d^4p}{(2\pi)^3} \eta_e(p.u)\left\{
\frac{-L^{(1)}_{i}\delta^\prime(p^2 - m^2)}{(p + q)^2 - m^2} +
\frac{L^{(2)}_{i}\delta(p^2 - m^2)}{[(p + q)^2 - m^2]^2}+ 
(q\rightarrow -q)\right\} \,,
\eeq
with
\beqa
L^{(1)}_{1} & = & (p_u + q_u)K +
p_b(p\cdot q + q^2) - m^2 q_b\,,\nonumber\\
L^{(2)}_{1} & = & -p_u K -
(p_b + q_b)p\cdot q + m^2 q_b \,,\nonumber\\
L^{(1)}_{2} & = & (p_b + q_b)K +
p_u(p\cdot q + q^2) - m^2 q_u \,,\nonumber\\
L^{(2)}_{2} & = &  -p_bK - 
(p_u + q_u)p\cdot q + m^2 q_u \,.
\eeqa
In writing the last equation we have introduced the factor
\beq
\label{K}
K = p_u q_b - p_b q_u\,,
\eeq
and used the shorthand notation
\beqa
\label{dotproducts1}
p_u & = & p\cdot u\,\nonumber\\
p_b & = & p\cdot b\,,
\eeqa
and similarly for $q_u$ and $q_b$.

\subsection{The structure of $T^{(A)}_{\mu\nu}$}

Since $T^{(A)}_{\mu\nu}$ satisfies $q^\nu T^{(A)}_{\mu\nu} = 0$
but it is not transverse in the index $\mu$, it can in principle
contain additional terms that are proportional to the tensors
$q^\mu \tilde t^\nu, q^\mu \tilde b^\nu, q^\mu \tilde u^\nu$.
However, an inspection of \Eq{TVAintegral} reveals that $T^{(A)}_{\mu\nu}$
is a true tensor rather than a pseudo-tensor and therefore the term
$q^\mu \tilde t^\nu$ cannot be present, and in addition that
there is no term proportional to $b_{\mu}b_{\nu}$, which is a consequence
of the calculation being up to linear terms in $B$.
Consequently $T^{(A)}_{\mu\nu}$ is of the form 
\beq
\label{TAform1loop}
T^{(A)}_{\mu\nu}  = T^{(A)}_L Q_{\mu\nu} + 
T^{(A)}_T \left(\tilde g_{\mu\nu} - Q_{\mu\nu}\right) + 
T^{(A)}_A A_{3\mu\nu} +
T^{(A)}_S S_{3\mu\nu} + 
T^{(A)}_u q_\mu \tilde u_\nu + 
T^{(A)}_b q_\mu \tilde b_\nu \,,
\eeq
where the integral formulas for the form factors are obtained 
by projecting \Eq{TVAintegral} with the tensors that appear in this expansion.
In this way, the form factors are given by
\beqa
\label{TAform}
T^{(A)}_X = 
4|e|B C_X\int\frac{d^4p}{(2\pi)^3} \eta_e(p)\left\{
\frac{-K^{(1)}_{X}\delta^\prime(p^2 - m^2)}{(p + q)^2 - m^2} +
\frac{K^{(2)}_{X}\delta(p^2 - m^2)}{[(p + q)^2 - m^2]^2}
+ (q\rightarrow -q)\right\} & \qquad & (X = L,T,S,A)\,,\nonumber\\
T^{(A)}_X = 
4|e|B C_X
\int\frac{d^4p}{(2\pi)^3} \eta_e(p)\left\{
\frac{-K^{(1)}_{X}\delta^\prime(p^2 - m^2)}{(p + q)^2 - m^2} +
\frac{K^{(2)}_{X}\delta(p^2 - m^2)}{[(p + q)^2 - m^2]^2}
- (q\rightarrow -q)\right\} & \qquad & (X = u,b)\,,
\eeqa
where
\beqa
\label{CX}
C_A = C_L = 2C_T & = & \frac{1}{\tilde u^2}\,,\nonumber\\
C_S & = & \left(\frac{1}{\tilde u^2\tilde b^2}\right)\,,\nonumber\\
C_u & = & \left(\frac{1}{q^2\tilde u^2}\right)\,,\nonumber\\
C_b & = & \left(\frac{1}{q^2\tilde b^2}\right)\,,
\eeqa
and
\beqa
\label{Kxy}
K^{(1)}_{L} & = & 2p_{\tilde u}K^\prime -
\tilde u^2 K\,,\nonumber\\
K^{(2)}_{L} & = & 2p_{\tilde u}\left(K^\prime - q_b\right) +
\tilde u^2 K\,,\nonumber\\
K^{(1)}_T & = & -2p_{\tilde u}K^\prime - 
\tilde u^2 K\,,\nonumber\\
K^{(2)}_T & = & -2p_{\tilde u}\left(K^\prime - q_b\right) + 
\tilde u^2 K\,,\nonumber\\
K^{(1)}_{S} & = & p_{\tilde b}K^\prime + 
p_{\tilde u}p_u\tilde b\cdot b\,,\nonumber\\
K^{(2)}_{S} & = & 
p_{\tilde b}\left(K^\prime - q_b\right) + 
p_{\tilde u}(p_u + q_u)\tilde b\cdot b\,,\nonumber\\
K^{(1)}_{u} & = & (p\cdot q + q^2)K^\prime
+ p_{\tilde u}K + m^2 q_b \nonumber\\
K^{(2)}_{u} & = & p\cdot q\left(K^\prime - q_b\right)
+ p_{\tilde u}K + m^2 q_b\,,\nonumber\\
K^{(1)}_{b} & = & p_{\tilde b}K + \tilde b\cdot b\left[
p_u(p\cdot q + q^2) - m^2 q_u\right]\,,\nonumber\\
K^{(2)}_{b} & = & p_{\tilde b}K + \tilde b\cdot b\left[
(p_u + q_u)p\cdot q - m^2 q_u\right]\,,\nonumber\\
K^{(1)}_{A} = K^{(2)}_{A} & = & -m^2\,,
\eeqa
where $K$ is defined in \Eq{K},
\beq
\label{kprime}
K^\prime = \tilde u\cdot b\,p_u - \tilde u^2\,p_b\,,
\eeq
and, in addition to \Eq{dotproducts1}, we have used the shorthand notation 
\beqa
\label{dotproducts2}
p_{\tilde b} & = & p\cdot\tilde b\,,\nonumber\\
p_{\tilde u} & = & p\cdot\tilde u\,.
\eeqa
In writing \Eq{Kxy} we have used the fact that 
$\tilde u^\mu$ and $\tilde b^\mu$ are orthogonal to $q^\mu$,
as well as the relations $u\cdot\tilde b = 0$ and 
$\tilde u^2\,\tilde b\cdot b = \tilde b^2$.

\subsection{Generic formulas}

The integrals involved to evaluate the form factors are of the
generic form
\beqa
I_1 & = & (-1)\int\frac{d^4p}{(2\pi)^3} \eta_e(p)
\frac{F_1(p,q)\delta^\prime(p^2 - m^2)}{(p + q)^2 - m^2} \,,\nonumber\\
I_2 & = & \int\frac{d^4p}{(2\pi)^3} \eta_e(p)
\frac{F_2(p,q)\delta(p^2 - m^2)}{[(p + q)^2 - m^2]^2} \,.
\eeqa
The various form factors differ only in the choice of the
functions $F_{1,2}$.
In $I_2$, the integration over $p^0$ can be carried out trivially,
and
\beq
\label{I2}
I_2 = \int\frac{d^3p}{(2\pi)^3 2E} \left.\left[
\frac{F_2(p,q)f_e}{[q^2 + 2p\cdot q]^2} +
\frac{F_2(-p,q)f_{\bar e}}{[q^2 - 2p\cdot q]^2}\right]
\right|_{p^0 = E(|\vec p|)} \,,
\eeq
where
\beq
\label{Ep}
E(p) = \sqrt{p^2 - m^2} \,.
\eeq
Using the relation
\beq
\partial_\mu\delta(p^2 - m^2) = 2p_\mu\delta^\prime(p^2 - m^2) \,,
\eeq
the variable $p^0$ can also be integrated out in $I_1$ by writing
\beq
\delta^\prime(p^2 - m^2) = 
\frac{1}{2p\cdot u}u^\mu\partial_\mu\delta(p^2 - m^2) \,,
\eeq
followed by an integration by parts. Thus,
\beqa
I_1 & = & u^\mu
\int\frac{d^4p}{(2\pi)^3}\delta(p^2 - m^2)\partial_\mu\left[
\left(\frac{1}{2p\cdot u}\right)\frac{\eta_e F_1(p,q)}{(p + q)^2 - m^2}\right]
\nonumber\\
& = & u^\mu\int\frac{d^3p}{(2\pi)^3 2E} \left.\left\{
\partial_\mu\left[\left(\frac{1}{2p\cdot u}\right)\left(
\frac{F_1(p,q)f_e}{(p + q)^2 - m^2} +
\frac{F_1(-p,q)f_{\bar e}}{(p - q)^2 - m^2}\right)\right]\right\}
\right|_{p^0 = E(|\vec p|)} \,.
\eeqa

In this fashion, we can write down the integral formulas for all
the form factors in terms of a pair of three-dimensional momentum 
integrals of the form of $I_{1,2}$, with the appropriate choices
of the functions $F_{1,2}$ that can be read-off directly from
\Eqs{TVform}{TAform}. 
There are no further manipulations that can be made to the integrals
$I_{1,2}$ that will allow us to evaluate the form factors in a general way. 
However, as we show below, the above formulas
are a useful starting point to compute explicitly
the form factors in various limiting cases.

%
% sec 5
%
\section{Approximate formulas}
\label{sec:approximate}

\subsection{Long wavelength limit}

A useful approximation, which is valid in many
practical situations, results from taking the so-called 
\emph{long wavelength limit}. The approximation is valid
in the regime in which the photon momentum $q\ll \langle\cal E\rangle$,
where $\langle\cal E\rangle$ denotes a typical average energy
of the particles in the background. 
The method to obtain the approximate formulas in this limit
is the same that was used in Ref.\ \cite{dnp1} and we will therefore
omit here some of the details. 

For the remainder of this section we set $u^\mu = (1,\vec 0)$, so that
all the kinematical variables refer to the medium rest frame as usual.
Let us consider first $I_2$. The method involves
making the substitutions $\vec p\rightarrow\vec p \mp \frac{1}{2}\vec Q$
in the two terms in the integrand of \Eq{I2}, respectively,
and then making a Taylor expansion in powers of $q/E$.
A useful auxiliary formula that is obtained as we have described is 
\beq
\label{Iaux}
\int\frac{d^3p}{(2\pi)^3}
\frac{{\cal F}(p)}{[q^2 + \lambda 2p\cdot q]^n}
= \lambda^n\int\frac{d^3p}{(2\pi)^3}
\frac{\left({\cal F} - \lambda\frac{\vec Q}{2}\cdot\frac{d{\cal F}}
{d\vec p}
- \lambda\frac{\omega}{E}{\cal F}\right)
\left(1 - \frac{\lambda n\omega}{E}\right)}
{[2E\omega - 2\vec p\cdot\vec Q)]^n}\,,
\eeq
where $p^0 = E(|\vec p|)$ and $\lambda = \pm 1$, and the total derivative
on the right-hand side is defined as
\beq
\frac{d}{d\vec p} = \frac{\partial}{\partial\vec p} + \frac{\vec p}{E}
\frac{\partial}{\partial E} \,.
\eeq

In order to write the resulting formula for $I_2$ concisely, it is useful to
define
\beq
\label{F2pm}
F^{\pm}_2(p,q) = \left.\left[\frac{f_e(p^0)F_2(p,q) \pm
f_{\bar e}(p^0)F_2(-p,q)}{2p^0}\right]\right|_{p^0 = E(|\vec p|)}\,.
\eeq
Then using \Eq{Iaux}, the procedure that we have outlined yields
\beq
\label{I2lw}
I_2 = \int\frac{d^3p}{(2\pi)^3}
\frac{F^{+}_2 - \frac{\vec Q}{2}\cdot\frac{d F^{-}_2}{d\vec p}
- \frac{\omega}{E}F^{-}_2}{[2E(\omega - \vec v\cdot\vec Q)]^2} \,,
\eeq
where
\beq
\vec v = \frac{\vec p}{E} \,.
\eeq
The result for $I_1$ is slightly more involved algebraically, but
straightforward to derive also.  In analogy with \Eq{F2pm} it is useful
to define
\beqa
\label{F1pm}
F^{\pm}_1(p,q) & = & \left.\left[\frac{f_e(p^0)F_1(p,q) \pm
f_{\bar e}(p^0)F_1(-p,q)}{2p^0}\right]\right|_{p^0 = E(p)}\,, \nonumber\\
F^{\prime\pm}_1(p,q) & = & \frac{1}{2E}
\frac{\partial F^{\pm}_1(p,q)}{\partial E}
\eeqa
in terms of which the long wave limit formula can be expressed
in the form
\beq
\label{I1lw}
I_1 = \int\frac{d^3p}{(2\pi)^3}\left\{
\frac{F^{\prime -}_1 - \frac{\vec Q}{2}\cdot\frac{d F^{\prime +}_1}{d\vec p}
- \frac{\omega}{2E}F^{\prime +}_1}{2E(\omega - \vec v\cdot\vec Q)} -
\frac{F^{+}_1 - \frac{\vec Q}{2}\cdot\frac{d F^{-}_1}{d\vec p}}
{[2E(\omega - \vec v\cdot\vec Q)]^2}\right\} \,.
\eeq
Since we are concerned only with the real part of the form factors,
the singularity of the integrand is to be handled by taking the
principal value part of the integral, as usual in these circumstances.

The formulas in \Eqs{I2lw}{I1lw} are particularly useful
for computing explicitly the form factors in various limiting cases,
as we now illustrate.

\subsection{Static limit and the neutrino induced charge}
\label{subsec:interpretation}
Similarly to the situation in the absence of the magnetic field\ \cite{np1},
it is possible to some extent to interpret some of the form factors,
or certain combinations of them, in terms
of electromagnetic moments of the neutrino. 
In order to identify the static electromagnetic moments
we have to look at the vertex function in the static limit
$\omega\rightarrow 0$ first, and then take the long wavelength limit
$Q\rightarrow 0$. 
It is well known that taking the limit in the
reverse order is not equivalent, since it represents a different
physical condition \cite{npzeromom}. 

However, in the present case there is the additional complication
that the vertex function, and whence the form factors, are not
isotropic functions of $\vec Q$ and instead they depend on the
components $Q_\parallel$ and $Q_\perp$ separately. Therefore,
the $Q\rightarrow 0$ limit can be taken in two different ways,
and in general the form factors will have different corresponding limit value.
In order to keep this clear, we use the notation
$\Gamma_\mu(\omega,Q_\perp,Q_\parallel)$ to denote the vertex 
function for any value of $q$, and 
$\Gamma_\mu(0,0,Q_\parallel\rightarrow 0)$ or
$\Gamma_\mu(0,Q_\perp\rightarrow 0,0)$ for its value
in the two limits that we have indicated. We use a similar notation for
the form factors.

If we insist in identifying a neutrino induced charge, or some
other induced electromagnetic moment for that matter, we are thus forced to 
define two different quantities, corresponding to the two different limits.
For example, the definition of the neutrino induced charge
used in Ref.\ \cite{np2} for the case of an isotropic medium,
must be amended as follows
\beqa
\label{eind}
e^{\parallel}_{\nu_\ell} = \frac{1}{2E_\nu}\mbox{Tr}\,
L\lslash{k}\Gamma^{(\ell)}_0(0,0,Q_\parallel\rightarrow 0) \,,\nonumber\\
e^{\perp}_{\nu_\ell} = \frac{1}{2E_\nu}\mbox{Tr}\,
L\lslash{k}\Gamma^{(\ell)}_0(0,Q_\perp\rightarrow 0,0) \,,
\eeqa
where $k^\mu = (E_\nu,\vec k)$ is the neutrino momentum vector.
For illustrative purposes we consider the evaluation specifically of the
form factors that enter in these two formulas.

We consider $e^{\parallel}_{\nu_\ell}$ first,
which corresponds to 
set $\vec Q_\perp = 0$, and then take the limit as $Q_\parallel\rightarrow 0$
last. As a practical trick, this amounts to evaluating the vertex function for
\beq
q^\mu = Q_\parallel b^\mu \,,
\eeq
and then take the $Q_\parallel\rightarrow 0$ limit. We therefore look at
\beq
\label{Gamma0staticparallel}
\Gamma^{(\ell)}_0(0,0,Q_\parallel) = 
\left(\frac{-eg^2 \chi^{(\ell)}_A}{4M^2_W}\right)
\left\{T^{(A)}_L\gamma_0 -
Q_\parallel T^{(A)}_u \vec\gamma\cdot\hat b\right\}L \,,
\eeq
where the form factors are evaluated at $\omega = 0$ and
$Q_\perp = 0$.
From the above formula for $e^{\parallel}_{\nu_\ell}$,
\Eq{Gamma0staticparallel} represents a contribution to that
quantity given by
\beq
\label{effparallel}
e^\parallel_{\nu_\ell} = \left(\frac{-eg^2 \chi^{(\ell)}_A}{4M_W^2}\right)
\lim_{Q_\parallel\rightarrow 0}\left.\left[
T^{(A)}_L - Q_\parallel T^{(A)}_u \hat k\cdot \hat b\right]
\right|_{\omega = 0,\;Q_\perp = 0}\,,
\eeq
where $\hat k$ is the neutrino momentum unit vector.

In order to evaluate $T^{(A)}_L$ and $T^{(A)}_u$ in this
limit we proceed as follows. First, from \Eq{Kxy},
for $q^\mu = Q_\parallel b^\mu$,
\beqa
K^{(2)}_{L} = K^{(1)}_{L}  &=  &
2p^0 P_\parallel + p^0 Q_\parallel\,,\nonumber\\
K^{(2)}_{u} = K^{(1)}_{u}  &=  &
(-Q_\parallel)\left[p^{0\,2} + P^2_\parallel + m^2 + Q_\parallel P_\parallel
\right] \,,
\eeqa
where we have decomposed
\beq
\vec p = \vec P_\perp + P_\parallel\hat b \,.
\eeq
For the particular kinematic
configuration that we are considering, the integrands do not depend explicitly
on $\vec P_\perp$, but only implicitly through $E(|\vec p|)$.
Therefore, the derivatives that appear in \Eq{F1pm}
can be rewritten using
\beq
\frac{1}{2E}\frac{\partial}{\partial E} = 
\frac{\partial}{\partial P^2_\perp} \,.
\eeq
Applying the results given in \Eqs{I2lw}{I1lw}, the integral formula
for $T^{(A)}_{L}$ becomes
\beqa
\label{TALstatic}
T^{(A)}_L(0,0,Q_\parallel) & = & (4|e|B)
\int\frac{d^2\vec P_\perp dP_\parallel}{(2\pi)^3}
\left(\frac{-1}{2P_\parallel}\right)
\frac{\partial}{\partial P_\perp^2}
\left[\left(1 - P_\parallel\frac{\partial}{\partial P_\parallel}\right)
(f_e + f_{\bar e})\right]\,,\nonumber\\
& = & 0
\eeqa
by symmetric integration over $P_\parallel$.
Similarly,
\beqa
T^{(A)}_u(0,0,Q_\parallel) & = & \frac{-1}{Q_\parallel}(4|e|B)
\int\frac{d^2\vec P_\perp dP_\parallel}{(2\pi)^3}
\frac{\partial}{\partial P^2_\perp}\left[\left(\frac{1}{2E^2} - 
\frac{1}{2P_\parallel}\frac{\partial}{\partial P_\parallel}\right)
E(f_e + f_{\bar e})\right] \,.
\eeqa
Putting $d^2\vec P_\perp = \pi dP^2_\perp$
the integral over the transverse components can be performed trivially, and
only the contribution from 
$P^2_\perp = 0$ survives. The remaining integral
over $P_\parallel$ can be written in a simple form yielding
\beq
\label{TAustatic}
T^{(A)}_u(0,0,Q_\parallel) = \frac{(4|e|B)}{Q_\parallel}t_\parallel \,,
\eeq
where
\beq
\label{tAu}
\label{tparallel}
t_\parallel = 
-\;\frac{1}{2}\int^\infty_{0}\frac{dp}{(2\pi)^2}
\left.\left[\frac{\partial}{\partial E} (f_e(E) + f_{\bar e}(E))
\right]\right|_{E = E(p)}\,.
\eeq
Using \Eqs{TALstatic}{TAustatic} in \Eq{effparallel}, we finally obtain
\beq
\label{effparallelfinal}
e^\parallel_{\nu_\ell} = \left(\frac{-e^2 g^2 \chi^{(\ell)}_A}{M^2_W}\right)
(\hat k\cdot\vec B)t_\parallel\,.
\eeq

The integral $t_\parallel$ cannot be evaluated in closed form
for an arbitrary distribution, but for particular cases it
is readily evaluated.

\subsubsection{Non-relativistic classical gas}
For example,
in the classical and non-relativistic limit we use 
\beq
\label{dfnr}
\frac{\partial f_{e,\bar e}}{\partial E}  = -\beta f_{e,\bar e}\,,
\eeq
where $\beta$ is the inverse temperature, and obtain
\beq
\label{TAustaticclassic}
t_\parallel = \frac{(n_e + n_{\bar e})\beta^2}{8m_e} \,.
\eeq

\subsubsection{Degenerate gas}
In this case, neglecting $f_{\bar e}$,
\beq
\label{TAutaticdeg}
t_\parallel = \left(\frac{E_F}{8\pi^2 p_F}\right) \,,
\eeq
where $p_F = (3\pi^2 n_e)^{1/3}$ is the Fermi momentum of the electron gas
and $E_F$ the corresponding energy. This result holds whether
the gas is relativistic or not.

A formula that resembles \Eq{effparallelfinal} was obtained in 
Ref.\ \cite{ganguly}. Although it was not explicitly stated there,
upon close inspection it becomes clear that the authors of that
reference took the $\vec Q\rightarrow 0$
limit by setting $Q_\perp = 0$ first, and then letting 
$Q_\parallel\rightarrow 0$ afterwards. Whence their calculation
corresponds to the calculation of the quantity that we have identified
as $e^\parallel_\nu$.
However, our result given in \Eq{effparallelfinal} differs
from the formula that is inferred from the results obtained 
in Ref.\ \cite{ganguly} [Eqs. (60) and (61)].
In fact, we can reproduce the result of that reference if we use
the relation given in \Eq{dfnr} to eliminate the derivatives
of the distribution functions in \Eq{tAu}.
However, \Eq{dfnr} is not valid for a general Fermi-Dirac distribution,
but only for the case of a non-relativistic and 
classical gas. For other cases,
the formula for the induced charge obtained in that reference cannot be used.
The appropriate formula for a degenerate gas is obtained
by using \Eq{TAutaticdeg} or, for an arbitrary distribution,
by using \Eq{tAu}.

A particular feature of \Eq{effparallelfinal}, which
was first noted in Ref.\ \cite{ganguly}, is the term
$\hat k\cdot \vec B$, which indicates a dependence of $e^\parallel_\nu$
on the direction of propagation of the neutrino.
It is important to keep in mind that this is a kinematic factor
that arises from taking the matrix element of the vertex function
between the neutrino spinors. The observation that we have made
above regarding the non-uniqueness of the zero momentum limit refers
to the dependence of the form factors themselves on
the variables $Q_\perp$ and $Q_\parallel$, independently of such kinematic
terms that may enter in specific matrix elements.

For completeness, we summarize below the results of the calculation
of $e^\perp_{\nu_\ell}$ which we have carried out following
a similar procedure. 

\subsection{The induced charge $e^\perp_{\nu_\ell}$}
For $\omega = 0$ and $Q_\parallel = 0$, we have
\beqa
T^{(V)}_{\mu 0} & = & 
\frac{1}{Q^2_\perp}T^{(V)}_2
i\epsilon_{\mu\alpha\beta\gamma}u^\alpha b^\beta q^\gamma \,,\nonumber\\
T^{(A)}_{\mu 0} & = & (T^{(A)}_S - T^{(A)}_A)b_\mu + T^{(L)}_A u_\mu
+ T^{(A)}_u q_\mu\,.
\eeqa
Either from the complete formulas given \Eq{TAform},
or from their long wavelength limit versions, it is easy to deduce that
\beq
T^{(A)}_L(0,Q_\perp,0) = T^{(A)}_u(0,Q_\perp,0) = 0\,,
\eeq
by focusing on the integration over $P_\parallel$ and observing
that in both cases the integrand is an odd function of that variable in this
configuration.
A less trivial result, but which follows straightforwardly
by using the formulas given in \Eqs{I2lw}{I1lw} is
\beq
T^{(V)}_2(0,Q_\perp\rightarrow 0,0) = Q^2_\perp \times\; \mbox{const}\,.
\eeq
Therefore,
\beq
\Gamma^{(\ell)}_0(0,Q_\perp\rightarrow 0,0) = \left(
\frac{-eg^2 \chi^{(\ell)}_A}{4M^2_W}\right)
\lim_{Q_\perp\rightarrow 0}\left.\left[
(T^{A)}_S - T^{(A)}_A)\lslash{b}L
\right]
\right|_{\omega = 0,\;Q_\parallel = 0}\,.
\eeq
Applying \Eqs{I2lw}{I1lw} once again to evaluate the
combination of form factors $T^{A)}_S - T^{(A)}_A$ we find
\beq
T^{(A)}_S(0,Q_\perp\rightarrow 0,0) - T^{(A)}_A(0,Q_\perp\rightarrow 0,0)
= 4|e|B t_\perp \,,
\eeq
where 
\beq
\label{tperp}
t_\perp = \int \frac{d^3p}{(2\pi)^3}\left.\left[\frac{1}{2E}
\frac{\partial}{\partial E}\left\{
\frac{1}{2E}\frac{\partial}{\partial E}\left(
\frac{(E^2 + P^2_\parallel + m^2)(f_e + f_{\bar e})}{2E}\right)
\right\}\right]\right|_{E = E(|\vec p|)} \,,
\eeq
and therefore, from \Eq{eind},
\beq
e^\perp_{\nu_\ell} = \left(\frac{-e^2 g^2\chi^{\ell)}_A}{M^2_W}\right)
(\hat k\cdot\vec B)t_\perp \,.
\eeq

For the case of a non-relativistic electron gas, \Eq{tperp} can
be manipulated to yield
\beq
\label{tperpnr}
t_\perp =
-\;\frac{1}{2}\int^\infty_{0}\frac{dp}{(2\pi)^2}
\left.\left[\frac{\partial}{\partial E} (f_e(E) + f_{\bar e}(E))
\right]\right|_{E = E(p)}\qquad(\mbox{non-relativistic case})\,,
\eeq
which is exactly the same as the integral in \Eq{tparallel}.
Therefore in this limit,
\beq
\label{unique}
e^\perp_{\nu_\ell} = e^\parallel_{\nu_\ell} 
\qquad(\mbox{non-relativistic case})\,.
\eeq
The result given in \Eq{tperpnr}, and consequently the equality between
$e^\parallel_{\nu_\ell}$ and $e^\parallel_{\nu_\ell}$,
is valid whether the gas is classical or degenerate, or in fact
for any Fermi-Dirac distribution consistent with the non-relativistic
limit. For a relativistic gas, the integrals $t_\perp$ and $t_\parallel$
are different, and \Eq{unique} no longer holds.

%
% sec 6
%
\section{Conclusions}
\label{sec:conclusions}
In this work we have been concerned with the electromagnetic
properties of a neutrino that propagates in a magnetized
electron background. Our goal has been to determine
the neutrino electromagnetic vertex function systematically, and in a way
that it be useful to study the neutrino processes
that may occur in such media, 
such as Cherenkov radiation and plasmon decay.

Our starting point was to use the electron propagator in 
the presence of a magnetic field to obtain 
the one-loop formula for the vertex function 
up to linear terms in the magnetic field.
We then decomposed the vertex function 
in terms of the minimal and complete set of tensors,
consistent with basic physical requirements such
as the transversality condition, and obtained the expressions for the
form factors in terms of a set of integrals over the momentum distribution
functions of the background electrons.
Simpler approximate formulas that are valid in the long wavelength 
limit, and which are useful for practical calculations of the form factors,
were given.

For illustrative purposes, and to make contact with previous
work, the calculation of the form factors that
enter in the effective neutrino charge was considered in some detail.
In connection with this we pointed out that, in contrast with
the situation in an isotropic medium (i.e., in the absence of the magnetic 
field), the static form factors do not have a unique value
in the zero (photon) momentum limit. We stress once more that this is not
a mathematical ambiguity, but it is actually a physical effect
that results from the fact that the medium in the present case
is essentially anisotropic. Furthermore, it is dynamical effect
quite different from the kinematic dependence that
the matrix elements of the vertex function may have on the
direction of propagation of the neutrino. Thus, for the
specific case of the neutrino induced charge, the two
quantities $e^\parallel_\nu$ and $e^\perp_\nu$ were introduced
and their expressions in terms of the form factors were given.
As we showed, they
have the same value for a non-relativistic gas, but are given
by different formulas otherwise.
In a given application all the form factors
are in principle relevant. 
The same method that we have used to
evaluate those that are related to the neutrino induced charge, 
can be used similarly to yield
the corresponding formulas for all the others.

In this work we have taken into
account only the electrons in the background. While the effects
of the nucleons are sometimes suppressed due to their mass, it
is known that in the presence of a magnetic field their 
contribution to some of the form factors is important.
The method and formulas that we have presented here provide
a firm basis, both conceptual and practical, to take those
effects into account as well. They provide a 
consistent basis to continue and extend the investigation
of problems in which the electromagnetic properties
of the neutrino and its coupling to a magnetic field
are believed to be important.

\acknowledgments
This material is based upon work supported  by the US National
Science Foundation under Grant No. 0139538.

\appendix
%
% appendix
%
\section{Proof of the transversality relations for $T^{(V)}_{\mu\nu}$}
\label{sec:transversalityTV}
We consider in some detail the proof of the relation
\beq
q^\nu T^{(V)}_{\mu\nu} = 0\,,
\eeq
where $T^{(V)}_{\mu\nu}$ is given in \Eq{TVAintegral}. Since
$T^{(V)}_{\mu\nu}$ is antisymmetric, this implies that
$q^\mu T^{(V)}_{\mu\nu} = 0$ also.

We write $T^{(V)}_{\mu\nu}$ in the following form
\beq
\label{piTB}
T^{(V)}_{\mu\nu} =
(4|e| B)\int\frac{d^4p}{(2\pi)^3} \eta_e(p) I^{(TB)}_{\mu\nu}
\eeq
where
\beq
I^{(TB)}_{\mu\nu} \equiv \left\{
\frac{-L^{(1)}_{\mu\nu}\delta^\prime(p^2 - m^2)}{d} +
\frac{L^{(2)}_{\mu\nu}\delta(p^2 - m^2)}{d^2}\right\}
- (q\rightarrow -q)\,.
\eeq
and
\beq
d \equiv p^{\prime\,2} - m^2 \,.
\eeq
Let us \emph{define} the quantity
\beq
\label{IT}
I^{(T)}_{\mu\nu} \equiv \delta(p^2 - m^2)\left[
\frac{L_{\mu\nu}}{d} + (q\rightarrow -q)\right] \,,
\eeq
where
\beq
\label{L}
L_{\mu\nu} = 2p_\mu p_\nu + p_\mu q\nu + q_\mu p_\nu - (p\cdot q)g_{\mu\nu}\,.
\eeq
We now state the following result, which is a purely algebraic
identity. Defining
\beq
P^{\mu\nu} \equiv i\epsilon^{\mu\nu\alpha\beta} b_\alpha u_\beta \,,
\eeq
then
\beq
\label{ITBrelation}
q^\mu I^{(TB)}_{\mu\nu} = 
\frac{i}{2}P^{\lambda\mu}\partial_\lambda I^{(T)}_{\mu\nu} \,,
\eeq
where $\partial_\lambda \equiv \partial/\partial p^\lambda$.
We prove \Eq{ITBrelation} below but, for the moment, notice that
it is all that we need. Observing that
\beq
P^{\lambda\mu}\partial_\lambda \eta(p) = 0
\eeq
(because $\partial_\lambda\eta \propto u_\lambda$), then
from \Eqs{piTB}{ITBrelation} we have
\beqa
q^\mu T^{(V)}_{\mu\nu} & = &
(4|e|B)\frac{i}{2}P^{\lambda\mu}\int\frac{d^4p}{(2\pi)^3}
\partial_\lambda  \left[\eta_e(p) I^{(T)}_{\mu\nu}\right]\nonumber\\
& = & 0 \,.
\eeqa

\subsection*{Proof of \Eq{ITBrelation}}

To prove \Eq{ITBrelation}, let us consider first its right-hand side,
\beq
P^{\lambda\mu}\partial_\lambda I^{(T)}_{\mu\nu} = 
P^{\lambda\mu}\left\{\partial_\lambda\left(
\frac{\delta(p^2 - m^2)L_{\mu\nu}}{d}\right) + 
(q\rightarrow -q)\right\} \,.
\eeq
Using
\beqa
\partial_\lambda \delta(p^2 - m^2) & = & 
\delta^\prime(p^2 - m^2)(2p_\lambda)\nonumber\\
\partial_\lambda\left(\frac{1}{d}\right) & = &
\frac{-2}{d^2}p^\prime_\lambda \nonumber\\
\partial_\lambda L_{\mu\nu} & = & 2g_{\mu\lambda}p_\nu + 2g_{\nu\lambda}p_\mu
+ g_{\mu\lambda}q_\nu + g_{\nu\lambda}q_\mu - g_{\mu\nu}q_\lambda \,,
\eeqa
we obtain
\beqa
P^{\lambda\mu}\partial_\lambda I^{(T)}_{\mu\nu} & = & 
\frac{2i\delta^\prime(p^2 - m^2)}{d}
\left[p_\nu\epsilon^{\mu\lambda\alpha\beta}
q_\mu p_\lambda u_\alpha b_\beta - (p\cdot q)\epsilon_{\nu\lambda\alpha\beta}
p^\lambda u^\alpha b^\beta\right]\nonumber\\
&&\mbox{} -
\frac{2i\delta(p^2 - m^2)}{d^2}\left[
p^\prime_\nu \epsilon^{\mu\lambda\alpha\beta} p_\mu q_\lambda u_\alpha b_\beta
- (p\cdot q)\epsilon_{\nu\lambda\alpha\beta}p^{\prime\lambda}u^\alpha b^\beta
\right]\nonumber\\
&&\mbox{}
- \frac{2i\delta(p^2 - m^2)}{d}\epsilon_{\nu\lambda\alpha\beta}
p^{\prime\lambda}u^\alpha b^\beta\nonumber\\
&&\mbox{} + (q\rightarrow - q) \,.
\eeqa
The third term can be rewritten thus,
\beq
\frac{\delta(p^2 - m^2)}{d} = \frac{\delta(p^2- m^2)}{d^2}(q^2 + 2p\cdot q)
\eeq
and then combining it with the second term, 
\beqa
\label{derIT}
P^{\lambda\mu}\partial_\lambda I^{(T)}_{\mu\nu} & = & 
\frac{2i\delta^\prime(p^2 - m^2)}{d}
\left[-p_\nu\epsilon^{\mu\lambda\alpha\beta}
p_\mu q_\lambda u_\alpha b_\beta - (p\cdot q)\epsilon_{\nu\lambda\alpha\beta}
p^\lambda u^\alpha b^\beta\right]\nonumber\\
&&\mbox{} -
\frac{2i\delta(p^2 - m^2)}{d^2}\left[
p^\prime_\nu \epsilon^{\mu\lambda\alpha\beta} p_\mu q_\lambda u_\alpha b_\beta
+ (p'\cdot q)\epsilon_{\nu\lambda\alpha\beta}p^{\prime\lambda}u^\alpha b^\beta
\right]\nonumber\\
&&\mbox{} + (q\rightarrow -q) \,.
\eeqa
This is all we will do with the right-hand side of \Eq{ITBrelation}. 

Now let us consider the left-hand side. From \Eq{L12},
we get
\beqa
\label{qL12}
q^\mu L^{(1)}_{\mu\nu} & = & -(p\cdot u)\epsilon_{\mu\nu\alpha\beta}
q^\mu p^\alpha b^\beta + (p\cdot b)\epsilon_{\mu\nu\alpha\beta}
q^\mu p^\alpha u^\beta + m^2\epsilon_{\mu\nu\alpha\beta}q^\mu u^\alpha b^\beta
\nonumber\\
q^\mu L^{(2)}_{\mu\nu} & = & (p'\cdot u)\epsilon_{\mu\nu\alpha\beta}
q^\mu p^\alpha b^\beta - (p'\cdot b)\epsilon_{\mu\nu\alpha\beta}
q^\mu p^\alpha u^\beta - m^2\epsilon_{\mu\nu\alpha\beta}q^\mu u^\alpha b^\beta
\nonumber\\
\eeqa
Using the identity given in \Eq{relacionpugnetera}, 
we obtain the following two identities,
\beqa
\label{identity1}
(p\cdot u)\epsilon_{\mu\nu\alpha\beta}q^\mu b^\alpha p^\beta -
(p\cdot b)\epsilon_{\mu\nu\alpha\beta}q^\mu u^\alpha p^\beta
+ p^2\epsilon_{\mu\nu\alpha\beta}q^\mu u^\alpha b^\beta =
\nonumber\\
-p_\nu\epsilon_{\mu\lambda\alpha\beta}p^\mu q^\lambda u^\alpha b^\beta
- (p\cdot q)\epsilon_{\nu\lambda\alpha\beta}p^\lambda u^\alpha b^\beta \,,
\eeqa
and
\beqa
\label{identity2}
(p'\cdot u)\epsilon_{\mu\nu\alpha\beta}q^\mu b^\alpha p^\beta -
(p'\cdot b)\epsilon_{\mu\nu\alpha\beta}q^\mu u^\alpha p^\beta
+ p^{\prime\,2}\epsilon_{\mu\nu\alpha\beta}q^\mu u^\alpha b^\beta =
\nonumber\\
-p^\prime_\nu\epsilon_{\mu\lambda\alpha\beta}p^\mu q^\lambda u^\alpha b^\beta
- (p'\cdot q)\epsilon_{\nu\lambda\alpha\beta}p^\lambda u^\alpha b^\beta \,,
\eeqa
and using them we can rewrite \Eq{qL12} in the form
\beqa
q^\mu L^{(1)}_{\mu\nu} & = &
-p_\nu\epsilon_{\mu\lambda\alpha\beta}p^\mu q^\lambda u^\alpha b^\beta
- (p\cdot q)\epsilon_{\nu\lambda\alpha\beta}p^\lambda u^\alpha b^\beta
- (p^2 - m^2)\epsilon_{\mu\nu\alpha\beta}q^\mu u^\alpha b^\beta \nonumber\\
q^\mu L^{(2)}_{\mu\nu} & = &
p'_\nu\epsilon_{\mu\lambda\alpha\beta}p^\mu q^\lambda u^\alpha b^\beta
+ (p'\cdot q)\epsilon_{\nu\lambda\alpha\beta}p^{\prime\lambda} u^\alpha b^\beta
+ (p^{\prime\,2} - m^2)\epsilon_{\mu\nu\alpha\beta}q^\mu u^\alpha b^\beta \,.
\nonumber\\
\eeqa
Therefore,
\beqa
\label{qITB}
q^\mu I^{(TB)}_{\mu\nu} & = & 
\left\{\frac{-q^\mu L^{(1)}_{\mu\nu}\delta^\prime(p^2 - m^2)}{d} +
\frac{q^\mu L^{(2)}_{\mu\nu}\delta(p^2 - m^2)}{d^2}\right\}
- (q\rightarrow -q)\nonumber\\
& = & -\frac{\delta^\prime(p^2 - m^2)}{d}\left[
-p_\nu\epsilon_{\mu\lambda\alpha\beta}p^\mu q^\lambda u^\alpha b^\beta
- (p\cdot q)\epsilon_{\nu\lambda\alpha\beta}p^\lambda u^\alpha b^\beta
- (p^2 - m^2)\epsilon_{\mu\nu\alpha\beta}q^\mu u^\alpha b^\beta\right] 
\nonumber\\
&&\mbox{} + \frac{\delta(p^2 - m^2)}{d^2}\left[
p'_\nu\epsilon_{\mu\lambda\alpha\beta}p^\mu q^\lambda u^\alpha b^\beta
+ (p'\cdot q)\epsilon_{\nu\lambda\alpha\beta}p^{\prime\lambda} u^\alpha b^\beta
+ (p^{\prime\,2} - m^2)\epsilon_{\mu\nu\alpha\beta}q^\mu u^\alpha b^\beta
\right]
\nonumber\\
&&\mbox{} + (q\rightarrow -q) \,.
\eeqa
The two terms that contain $(p^2 - m^2)$ and $(p^{\prime\,2} - m^2)$
combine to give
\beq
\frac{1}{d}\left[
\delta^\prime(p^2 - m^2)(p^2 - m^2) + \delta(p^2 - m^2)\right] = 0\,,
\eeq
where we have used the relation 
\beq
\label{deltaprime}
x\delta^\prime(x) = -\delta(x) \,.
\eeq
The remaining terms precisely coincide with \Eq{derIT}, which
therefore proves \Eq{ITBrelation}.

%
% appendix
%
\section{Proof of the transversality relations for $T^{(A)}_{\mu\nu}$}
\label{sec:transversalityTA}

The relation
\beq
q^\nu T^{(A)}_{\mu\nu} = 0\,,
\eeq
is proven similarly. From \Eq{TVAintegral},
\beq
\label{qTA}
q^\nu T^{(A)}_{\mu\nu} = 
(4|e|B)\int\frac{d^4p}{(2\pi)^3} \eta_e(p)\left\{
\frac{-q^\nu K^{(1)}_{\mu\nu}\delta^\prime(p^2 - m^2)}{(p + q)^2 - m^2} +
\frac{q^\nu K^{(2)}_{\mu\nu}\delta(p^2 - m^2)}{[(p + q)^2 - m^2]^2}
- (q\rightarrow -q)\right\}\,,
\eeq
where, from the defining equation \Eq{K12}, we obtain
\beqa
q^\nu K^{(1)}_{\mu\nu} & = & p_\mu\left[
-(p\cdot b)(u\cdot q) + (p\cdot u)(b\cdot q)\right] +
u_\mu\left[-(p\cdot b)(p^\prime\cdot q) - m^2 b\cdot q\right] +
b_\mu\left[(p\cdot u)(p^\prime\cdot q) + m^2 u\cdot q\right]\,,\nonumber\\
q^\nu K^{(2)}_{\mu\nu} & = & p^\prime_\mu\left[
-(p\cdot b)(u\cdot q) + (p\cdot u)(b\cdot q)\right] +
u_\mu\left[-(p^\prime\cdot b)(p\cdot q) - m^2 b\cdot q\right] +
b_\mu\left[(p^\prime\cdot u)(p\cdot q) + m^2 u\cdot q\right]\,.
\eeqa
Then, in analogy with \Eq{IT}, we define
\beq
\label{JT}
J^{(T)}_{\mu\nu} = \delta(p^2 - m^2)
\frac{\epsilon_{\mu\nu\alpha\beta}p^\alpha q^\beta}{d}\,,
\eeq
and by straightforward calculation of 
$\partial_\lambda J^{(T)}_{\mu\nu}$ it follows that
\beqa
\label{qK}
\frac{-q^\nu K^{(1)}_{\mu\nu}}{d}\delta^\prime(p^2 - m^2)
+ \frac{q^\nu K^{(2)}_{\mu\nu}}{d^2}\delta(p^2 - m^2) & = &
\left(\frac{-1}{2}\right)\left[
\frac{1}{i}P^{\nu\lambda}\partial_\lambda J^{(T)}_{\mu\nu} +
\frac{\delta(p^2 - m^2)}{d} (2Y_\mu)\right]\nonumber\\ 
&&\mbox{} +
\left[X_\mu(q^2 + 2p\cdot q) + Y_\mu(p^2 - m^2)\right]
\left[\frac{\delta(p^2 - m^2)}{d^2} - 
\frac{\delta^\prime(p^2 - m^2)}{d}\right]\nonumber\\
& = & 
\left(\frac{i}{2}\right)P^{\nu\lambda}\partial_\lambda J^{(T)}_{\mu\nu} -
X_\mu\delta^\prime(p^2 - m^2) \,.
\eeqa
In \Eq{qK} we have defined
\beqa
X_\mu & = & b_\mu(u\cdot p) - u_\mu(b\cdot p) \,,\nonumber\\
Y_\mu & = & u_\mu(b\cdot q) - b_\mu(u\cdot q) \,,
\eeqa
and to arrive at the last equality we have used the relation
of \Eq{deltaprime} once more.
Thus, using \Eq{qK} in \Eq{qTA},
\beqa
q^\nu T^{(A)}_{\mu\nu} & = &
(4|e|B)\frac{i}{2}P^{\nu\lambda}\int\frac{d^4p}{(2\pi)^3}
\partial_\lambda \left[\eta_e(p) J^{(T)}_{\mu\nu} - (q\rightarrow -q)\right]
\nonumber\\
& = & 0 \,.
\eeqa

On the other hand, by separating the tensors $K^{(1,2)}_{\mu\nu}$ into
their symmetric and antisymmetric parts, it is easy to show that
\beq
q^\mu T^{(A)}_{\mu\nu} = (4|e|B)(4m^2 Y_\nu)
\int\frac{d^4 p}{(2\pi)^3}\eta_e(p)\left[\frac{\delta(p^2 - m^2)}{d^2}
+ (q\rightarrow -q)\right] \,,
\eeq
which explicitly verifies that, as expected, 
$T^{(A)}_{\mu\nu}$ is not transverse with respect to its first index.
In this respect, the tensor $T^{(A)}_{\mu\nu}$, calculated in
the presence of the magnetic field as we re doing here, differs from 
the corresponding quantity calculated without the magnetic field,
in which case it is transverse in the first index as well.


\begin{thebibliography}{24}
\expandafter\ifx\csname natexlab\endcsname\relax\def\natexlab#1{#1}\fi
\expandafter\ifx\csname bibnamefont\endcsname\relax
  \def\bibnamefont#1{#1}\fi
\expandafter\ifx\csname bibfnamefont\endcsname\relax
  \def\bibfnamefont#1{#1}\fi
\expandafter\ifx\csname citenamefont\endcsname\relax
  \def\citenamefont#1{#1}\fi
\expandafter\ifx\csname url\endcsname\relax
  \def\url#1{\texttt{#1}}\fi
\expandafter\ifx\csname urlprefix\endcsname\relax\def\urlprefix{URL }\fi
\providecommand{\bibinfo}[2]{#2}
\providecommand{\eprint}[2][]{\url{#2}}

\bibitem[{\citenamefont{Nieves and Pal}(1989)}]{np1}
\bibinfo{author}{\bibfnamefont{J.~F.} \bibnamefont{Nieves}} \bibnamefont{and}
  \bibinfo{author}{\bibfnamefont{P.~B.} \bibnamefont{Pal}},
  \bibinfo{journal}{Phys. Rev. D} \textbf{\bibinfo{volume}{40}},
  \bibinfo{pages}{1693} (\bibinfo{year}{1989}).

\bibitem[{\citenamefont{Oraevsky et~al.}(1986)\citenamefont{Oraevsky, Semikoz,
  and Smorodinsky}}]{oraevsky}
\bibinfo{author}{\bibfnamefont{V.~N.} \bibnamefont{Oraevsky}},
  \bibinfo{author}{\bibfnamefont{V.~B.} \bibnamefont{Semikoz}},
  \bibnamefont{and} \bibinfo{author}{\bibfnamefont{Y.~A.}
  \bibnamefont{Smorodinsky}}, \bibinfo{journal}{Sov. Phys. JETP Lett.}
  \textbf{\bibinfo{volume}{43}}, \bibinfo{pages}{709} (\bibinfo{year}{1986}).

\bibitem[{\citenamefont{C.D'Olivo et~al.}(1989)\citenamefont{C.D'Olivo, Nieves,
  and Pal}}]{dnp1}
\bibinfo{author}{\bibfnamefont{J.}~\bibnamefont{C.D'Olivo}},
  \bibinfo{author}{\bibfnamefont{J.~F.} \bibnamefont{Nieves}},
  \bibnamefont{and} \bibinfo{author}{\bibfnamefont{P.~B.} \bibnamefont{Pal}},
  \bibinfo{journal}{Phys. Rev. D} \textbf{\bibinfo{volume}{40}},
  \bibinfo{pages}{3679} (\bibinfo{year}{1989}).

\bibitem[{\citenamefont{Sawyer}(1992)}]{sawyer}
\bibinfo{author}{\bibfnamefont{R.~F.} \bibnamefont{Sawyer}},
  \bibinfo{journal}{Phys. Rev. D} \textbf{\bibinfo{volume}{46}},
  \bibinfo{pages}{1180} (\bibinfo{year}{1992}).

\bibitem[{\citenamefont{D'Olivo et~al.}(1990)\citenamefont{D'Olivo, Nieves, and
  Pal}}]{dnp2}
\bibinfo{author}{\bibfnamefont{J.~C.} \bibnamefont{D'Olivo}},
  \bibinfo{author}{\bibfnamefont{J.~F.} \bibnamefont{Nieves}},
  \bibnamefont{and} \bibinfo{author}{\bibfnamefont{P.~B.} \bibnamefont{Pal}},
  \bibinfo{journal}{Phys. Rev. Lett.} \textbf{\bibinfo{volume}{64}},
  \bibinfo{pages}{1088} (\bibinfo{year}{1990}).

\bibitem[{\citenamefont{Semikoz and Valle}(1994)}]{semikoz}
\bibinfo{author}{\bibfnamefont{V.~B.} \bibnamefont{Semikoz}} \bibnamefont{and}
  \bibinfo{author}{\bibfnamefont{J.~W.~F.} \bibnamefont{Valle}},
  \bibinfo{journal}{Nucl. Phys. B} \textbf{\bibinfo{volume}{425}},
  \bibinfo{pages}{65} (\bibinfo{year}{1994}).

\bibitem[{\citenamefont{Semikoz and Valle}(1997)}]{semikozE}
\bibinfo{author}{\bibfnamefont{V.~B.} \bibnamefont{Semikoz}} \bibnamefont{and}
  \bibinfo{author}{\bibfnamefont{J.~W.~F.} \bibnamefont{Valle}},
  \bibinfo{journal}{Nucl. Phys. B} \textbf{\bibinfo{volume}{485}},
  \bibinfo{pages}{585} (\bibinfo{year}{1997}).

\bibitem[{\citenamefont{D'Olivo and Nieves}(1996)}]{dn1}
\bibinfo{author}{\bibfnamefont{J.~C.} \bibnamefont{D'Olivo}} \bibnamefont{and}
  \bibinfo{author}{\bibfnamefont{J.~F.} \bibnamefont{Nieves}},
  \bibinfo{journal}{Phys. Lett. B} \textbf{\bibinfo{volume}{383}},
  \bibinfo{pages}{87} (\bibinfo{year}{1996}).

\bibitem[{\citenamefont{D'Olivo et~al.}(1996)\citenamefont{D'Olivo, Nieves, and
  Pal}}]{dnp3}
\bibinfo{author}{\bibfnamefont{J.~C.} \bibnamefont{D'Olivo}},
  \bibinfo{author}{\bibfnamefont{J.~F.} \bibnamefont{Nieves}},
  \bibnamefont{and} \bibinfo{author}{\bibfnamefont{P.~B.} \bibnamefont{Pal}},
  \bibinfo{journal}{Phys. Lett. B} \textbf{\bibinfo{volume}{365}},
  \bibinfo{pages}{178} (\bibinfo{year}{1996}).

\bibitem[{\citenamefont{Kusenko and Segr\'e}(1996)}]{kicks1}
\bibinfo{author}{\bibfnamefont{A.}~\bibnamefont{Kusenko}} \bibnamefont{and}
  \bibinfo{author}{\bibfnamefont{G.}~\bibnamefont{Segr\'e}},
  \bibinfo{journal}{Phys. Rev. Lett.} \textbf{\bibinfo{volume}{77}},
  \bibinfo{pages}{4872} (\bibinfo{year}{1996}).

\bibitem[{\citenamefont{Kusenko and Segr\'e}(1997{\natexlab{a}})}]{kicks2}
\bibinfo{author}{\bibfnamefont{A.}~\bibnamefont{Kusenko}} \bibnamefont{and}
  \bibinfo{author}{\bibfnamefont{G.}~\bibnamefont{Segr\'e}},
  \bibinfo{journal}{Phys. Lett. B} \textbf{\bibinfo{volume}{396}},
  \bibinfo{pages}{197} (\bibinfo{year}{1997}{\natexlab{a}}).

\bibitem[{\citenamefont{Kusenko and Segr\'e}(1999)}]{kicks3}
\bibinfo{author}{\bibfnamefont{A.}~\bibnamefont{Kusenko}} \bibnamefont{and}
  \bibinfo{author}{\bibfnamefont{G.}~\bibnamefont{Segr\'e}},
  \bibinfo{journal}{Phys. Rev. D} \textbf{\bibinfo{volume}{59}},
  \bibinfo{pages}{061302} (\bibinfo{year}{1999}).

\bibitem[{\citenamefont{Kusenko and Segr\'e}(1997{\natexlab{b}})}]{kicks4}
\bibinfo{author}{\bibfnamefont{A.}~\bibnamefont{Kusenko}} \bibnamefont{and}
  \bibinfo{author}{\bibfnamefont{G.}~\bibnamefont{Segr\'e}},
  \bibinfo{journal}{Phys. Rev. Lett.} \textbf{\bibinfo{volume}{79}},
  \bibinfo{pages}{2751} (\bibinfo{year}{1997}{\natexlab{b}}).

\bibitem[{\citenamefont{Qian}(1997)}]{kicks5}
\bibinfo{author}{\bibfnamefont{Y.~Z.} \bibnamefont{Qian}},
  \bibinfo{journal}{Phys. Rev. Lett.} \textbf{\bibinfo{volume}{79}},
  \bibinfo{pages}{2750} (\bibinfo{year}{1997}).

\bibitem[{\citenamefont{Nunokawa et~al.}(1997)\citenamefont{Nunokawa, Semikoz,
  Smirnov, and Valle}}]{nunokawa}
\bibinfo{author}{\bibfnamefont{H.}~\bibnamefont{Nunokawa}},
  \bibinfo{author}{\bibfnamefont{V.~B.} \bibnamefont{Semikoz}},
  \bibinfo{author}{\bibfnamefont{A.~Y.} \bibnamefont{Smirnov}},
  \bibnamefont{and} \bibinfo{author}{\bibfnamefont{J.~W.~F.}
  \bibnamefont{Valle}}, \bibinfo{journal}{Nucl. Phys. B}
  \textbf{\bibinfo{volume}{501}}, \bibinfo{pages}{17} (\bibinfo{year}{1997}).

\bibitem[{\citenamefont{Esposito and Capone}(1996)}]{esposito}
\bibinfo{author}{\bibfnamefont{S.}~\bibnamefont{Esposito}} \bibnamefont{and}
  \bibinfo{author}{\bibfnamefont{G.}~\bibnamefont{Capone}},
  \bibinfo{journal}{Z. Phys. \textbf{C}} \textbf{\bibinfo{volume}{70}},
  \bibinfo{pages}{55} (\bibinfo{year}{1996}).

\bibitem[{\citenamefont{Erdas et~al.}(1998)\citenamefont{Erdas, Kim, and
  Lee}}]{kimetal}
\bibinfo{author}{\bibfnamefont{A.}~\bibnamefont{Erdas}},
  \bibinfo{author}{\bibfnamefont{C.~W.} \bibnamefont{Kim}}, \bibnamefont{and}
  \bibinfo{author}{\bibfnamefont{T.~H.} \bibnamefont{Lee}},
  \bibinfo{journal}{Phys. Rev. D} \textbf{\bibinfo{volume}{58}},
  \bibinfo{pages}{08516} (\bibinfo{year}{1998}).

\bibitem[{\citenamefont{Elmfors et~al.}(1996)\citenamefont{Elmfors, Grasso, and
  Raffelt}}]{elmfors}
\bibinfo{author}{\bibfnamefont{P.}~\bibnamefont{Elmfors}},
  \bibinfo{author}{\bibfnamefont{D.}~\bibnamefont{Grasso}}, \bibnamefont{and}
  \bibinfo{author}{\bibfnamefont{G.}~\bibnamefont{Raffelt}},
  \bibinfo{journal}{Nucl. Phys. B} \textbf{\bibinfo{volume}{479}},
  \bibinfo{pages}{3} (\bibinfo{year}{1996}).

\bibitem[{\citenamefont{D'Olivo and Nieves}(1997)}]{dn2}
\bibinfo{author}{\bibfnamefont{J.~C.} \bibnamefont{D'Olivo}} \bibnamefont{and}
  \bibinfo{author}{\bibfnamefont{J.~F.} \bibnamefont{Nieves}},
  \bibinfo{journal}{Phys. Rev. D} \textbf{\bibinfo{volume}{56}},
  \bibinfo{pages}{5898} (\bibinfo{year}{1997}).

\bibitem[{\citenamefont{Ioannisian and Raffelt}(1997)}]{raffelt}
\bibinfo{author}{\bibfnamefont{A.~N.} \bibnamefont{Ioannisian}}
  \bibnamefont{and} \bibinfo{author}{\bibfnamefont{G.~G.}
  \bibnamefont{Raffelt}}, \bibinfo{journal}{Phys. Rev. D}
  \textbf{\bibinfo{volume}{55}}, \bibinfo{pages}{7038} (\bibinfo{year}{1997}).

\bibitem[{\citenamefont{Bhattacharya et~al.}(2001)\citenamefont{Bhattacharya,
  Ganguly, and Konar}}]{ganguly}
\bibinfo{author}{\bibfnamefont{K.}~\bibnamefont{Bhattacharya}},
  \bibinfo{author}{\bibfnamefont{A.}~\bibnamefont{Ganguly}}, \bibnamefont{and}
  \bibinfo{author}{\bibfnamefont{S.}~\bibnamefont{Konar}},
  \bibinfo{journal}{Phys. Rev. D} \textbf{\bibinfo{volume}{65}},
  \bibinfo{pages}{013007} (\bibinfo{year}{2001}).

\bibitem[{\citenamefont{D'Olivo et~al.}(2003)\citenamefont{D'Olivo, Nieves, and
  Sahu}}]{dns}
\bibinfo{author}{\bibfnamefont{J.~C.} \bibnamefont{D'Olivo}},
  \bibinfo{author}{\bibfnamefont{J.~F.} \bibnamefont{Nieves}},
  \bibnamefont{and} \bibinfo{author}{\bibfnamefont{S.}~\bibnamefont{Sahu}},
  \bibinfo{journal}{Phys. Rev. D} \textbf{\bibinfo{volume}{67}},
  \bibinfo{pages}{025018} (\bibinfo{year}{2003}).

\bibitem[{\citenamefont{Nieves and Pal}(1995)}]{npzeromom}
\bibinfo{author}{\bibfnamefont{J.~F.} \bibnamefont{Nieves}} \bibnamefont{and}
  \bibinfo{author}{\bibfnamefont{P.~B.} \bibnamefont{Pal}},
  \bibinfo{journal}{Phys. Rev. D} \textbf{\bibinfo{volume}{51}},
  \bibinfo{pages}{5300} (\bibinfo{year}{1995}), \bibinfo{note}{and references
  there in.}

\bibitem[{\citenamefont{Nieves and Pal}(1994)}]{np2}
\bibinfo{author}{\bibfnamefont{J.~F.} \bibnamefont{Nieves}} \bibnamefont{and}
  \bibinfo{author}{\bibfnamefont{P.~B.} \bibnamefont{Pal}},
  \bibinfo{journal}{Phys. Rev. D} \textbf{\bibinfo{volume}{49}},
  \bibinfo{pages}{1398} (\bibinfo{year}{1994}).

\end{thebibliography}
\end{document}